\documentclass[a4paper,superscriptaddress,floatfix]{revtex4}

\usepackage[normalem]{ulem}

\usepackage{epsfig}
\usepackage{amsfonts}
\usepackage{amsmath}
\usepackage{amssymb}
\usepackage{amsthm}
\usepackage{float}
\usepackage{array}
\usepackage{booktabs}
\usepackage{color}
\usepackage{graphicx}
\usepackage{hyperref}
\usepackage{multirow}
\usepackage{slashed}
\usepackage{subfigure}
\usepackage{theorem}
\usepackage{textcomp}
\usepackage[utf8]{inputenc}

\allowdisplaybreaks[1]

\numberwithin{equation}{section}

\begin{document}

\begin{flushright}
MS-TP-17-34\\
RAL-P-2018-001
\end{flushright}

\title{Constraining parton distribution functions \\
from neutral current Drell-Yan measurements}

\author{E. Accomando}
\email[E-mail: ]{e.accomando@soton.ac.uk}
\affiliation{School of Physics \& Astronomy, University of Southampton,
        Highfield, Southampton SO17 1BJ, United Kingdom}
\affiliation{Particle Physics Department, Rutherford Appleton Laboratory, 
       Chilton, Didcot, Oxon OX11 0QX, United Kingdom}

\author{J. Fiaschi}
\email[E-mail: ]{juri.fiaschi@soton.ac.uk}
\affiliation{School of Physics \& Astronomy, University of Southampton,
        Highfield, Southampton SO17 1BJ, United Kingdom}
\affiliation{Particle Physics Department, Rutherford Appleton Laboratory, 
       Chilton, Didcot, Oxon OX11 0QX, United Kingdom}
\affiliation{Institut f{\" u}r Theoretische Physik, 
Universit{\" a}t M{\" u}nster, D 48149 M{\" u}nster, 
Germany}

\author{F. Hautmann}
\email[E-mail: ]{hautmann@thphys.ox.ac.uk}
\affiliation{Particle Physics Department, Rutherford Appleton Laboratory, 
       Chilton, Didcot, Oxon OX11 0QX, United Kingdom}
\affiliation{Elementaire Deeltjes Fysica, Universiteit Antwerpen, B 2020 Antwerpen, Belgium}
\affiliation{Theoretical Physics Department, University of Oxford, Oxford OX1 3NP, United Kingdom}

\author{S. Moretti}
\email[E-mail: ]{s.moretti@soton.ac.uk}
\affiliation{School of Physics \& Astronomy, University of Southampton,
        Highfield, Southampton SO17 1BJ, United Kingdom}
\affiliation{Particle Physics Department, Rutherford Appleton Laboratory, 
       Chilton, Didcot, Oxon OX11 0QX, United Kingdom}

 \begin{abstract}
{
We study the cross section $\sigma$ and Forward-Backward asymmetry ($A_{\rm FB}$) in the process $pp\to \gamma^*,Z\to \ell^+\ell^-$ (with $\ell=e,\mu$) for determinations of Parton Distribution Functions (PDFs) of the proton.
We show that, once mapped in the invariant mass of the di-lepton final state, $M({\ell\ell})$, both observables, $\sigma$ and $A_{\rm FB}$, display a statistical error
which is presently competitive with that assigned to the existing PDF sets and which will rapidly become smaller than the latter as the luminosity being accumulated at Run-II of the LHC grows.
This statement is applicable to both on-peak and off-peak $M({\ell\ell})$ regions, both (just) below and above it, thereby offering a means of constraining the quark PDFs over a sizeable $(x,Q^2)$ range.}

\end{abstract}

\maketitle

\setcounter{footnote}{0}

\section{Introduction}

Neutral Current (NC) and Charged Current (CC) Drell-Yan (DY) production modes in hadronic collisions are used both for searching for Beyond the Standard Model (BSM) physics and for testing the Standard Model.
In the first case, they are the cleanest processes where extra gauge bosons could appear (see for instance Refs.~\cite{Langacker:2008yv,Accomando:2010fz,Accomando:2011eu,Accomando:2013sfa,Accomando:2015cfa,Accomando:2016sge,Accomando:2017fmb}
for a sample of works on this subject); in the second case, they are extremely valuable as they allow one to access the structure of the (anti)quark dynamics inside the (anti)proton.
In this paper, we will be focusing on this latter feature.
This approach has been exploited consistently over the course of decades, from the pioneer S$p\bar p$S collider (at CERN) to the $p\bar p$ Tevatron (at FNAL) and the presently running $pp$ LHC (at CERN). 
In the CC mode the total cross section $\sigma$ is significantly larger than in the NC mode.
Differential cross sections have been investigated for both CC and NC modes to constrain Parton Distribution Functions (PDFs).
See for instance~\cite{Alekhin:2017olj,Alekhin:2017kpj,Alekhin:2015cza,Ball:2017nwa,Dulat:2015mca,Aaboud:2016btc} for recent studies.

As far as asymmetries go, the following one defined in the CC production, 
\begin{equation}
\label{AFB-CC}
 A_{\rm CC} = \frac { d \sigma / d \eta(\ell^+) - d \sigma / d \eta(\ell^-) }  
                                                 { d \sigma / d \eta(\ell^+) + d \sigma / d \eta(\ell^-) }, 
\end{equation}
has traditionally been used to constrain PDFs, particularly the ratio of $d$-quark to $u$-quark distributions. 
This is called the `lepton charge asymmetry' and is expressed in terms of the lepton pseudo-rapidity, $\eta({\ell^\pm})$, where the charged lepton $\ell^\pm$ comes from the decay of a $W^\pm$-boson with the same charge.
Just like any asymmetry, this one in the $W^\pm$-boson mediated channel is a powerful tool for PDF fitting.
The charge asymmetry is in fact a ratio of cross sections in which many systematical errors are cancelled out.
This leads to a greater precision, especially when extracting information on the $d/u$ PDF ratio, albeit at a price of a statistical error propagation from the cross sections. 

The $A_{\rm CC}$ observable is zero for the case of the $\gamma^*,Z$ mediated channel.
In this case, however, one can define the Forward-Backward (FB) asymmetry
\begin{equation}
\label{AFB-NC}
  A_{\rm NC} = \frac { d \sigma / d M(\ell^+\ell^-)[\eta(\ell^+)>0] - d \sigma / d M(\ell^+\ell^-)[\eta(\ell^+)<0] }  
                                 { d \sigma / d M(\ell^+\ell^-)[\eta(\ell^+)>0] + d \sigma / d M(\ell^+\ell^-)[\eta(\ell^+)<0] },
\end{equation}
where the identification of the Forward (F) and Backward (B) hemispheres via the $\eta(\ell^+)>0$ and $\eta(\ell^+)<0$ restrictions is obtained through the exploitation of the
di-lepton boost variable~\cite{Dittmar:1996my,Accomando:2015cfa,Accomando:2016tah,Accomando:2016ehi}.
Hence, this is effectively a ``di-lepton FB asymmetry" (henceforth, denoted as $A_{\rm FB}$).

The advantage of the FB asymmetry is that one can define it both on-shell (i.e., for $M({\ell^+\ell^-}) = M_Z$) and off-shell (i.e., for $M({\ell^+\ell^-})\ne M_Z$), while in the CC case it is not possible to define
these two regions (because of the escaping neutrino preventing the reconstruction of the $W^\pm$-boson mass).
Hence, in the latter case, the peak region necessarily dominates, so that the $(x,Q^2)$ ranges testable for the (anti)quark PDFs are necessarily reduced with respect to the ones accessible via the former
(where the off-peak region can efficiently be selected).
Conversely, for the FB asymmetry, access to the $d/u$ PDF ratio is more indirect than for the case of the CC one.
The CC and NC modes have thus complementary strengths.

The impact of CC and NC doubly differential measurements in the di-lepton mass and rapidity has recently been examined, see e.g.~\cite{Alekhin:2017olj,Ball:2017nwa,Aaboud:2016btc}. 
It is the purpose of this paper, by leveraging on the fact that, presently, there exist $\sigma$ and $A_{\rm FB}$ measurements in the NC mode at the LHC in the aforementioned regions of the di-lepton mass spectrum,
to investigate the role of the angular information encoded in the $A_{\rm FB}$, which is related to the single-lepton pseudorapidity.
This information, once combined with di-lepton mass and rapidity, would qualitatively correspond to triple differential cross sections. These cross sections, classified as either forward or backward, can be in fact used to reconstruct an experimental measurement of $A_{\rm FB}$, differentially in the invariant mass and rapidity of the di-lepton system.
Measurements of the Drell-Yan triple-differential cross sections have been performed for electron and muon pairs. The ATLAS analysis ~\cite{Aaboud:2017ffb} uses 20.2 fb$^{-1}$ of $pp$ collision data at $\sqrt{s}$ = 8 TeV. These precision data are being used for the determination of the effective weak mixing angle. They are also sensitive to the parton distribution functions. However, they have not yet been used systematically for PDF determinations.
The results of our investigation show that the statistical error reachable at the LHC Run-II can be lower than the one (current and foreseen) quoted for the (anti)quark PDFs by various groups.
On one hand, while benefiting from the large statistics around the $Z$-boson peak, the FB asymmetry allows one to probe the large-$x$ domain, where the PDF error is largest, if a cut on the rapidity of the di-lepton system is implemented.
On the other hand, we herald the message that $Z$-boson cross section and FB asymmetry measurements, especially off-peak, can help in PDF determinations over a significant $Q^2$ range.
We will illustrate our results by using two specific up-to-date PDF sets, CT14NNLO~\cite{Dulat:2015mca} and NNPDF3.1~\cite{Ball:2017nwa}. 

\begin{figure}[t]
\begin{center}
\includegraphics[width=0.47\textwidth]{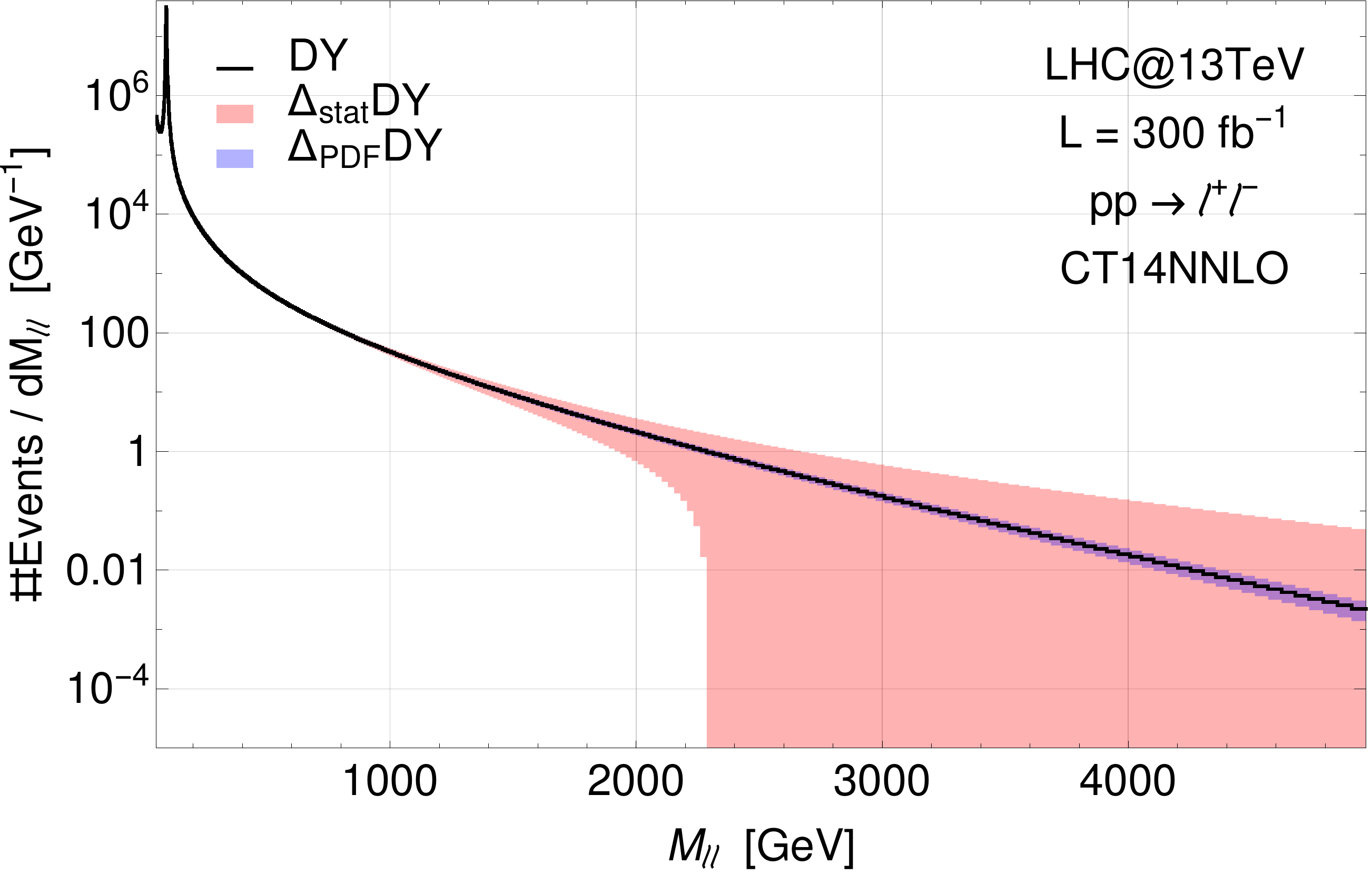}{(a)}
\includegraphics[width=0.47\textwidth]{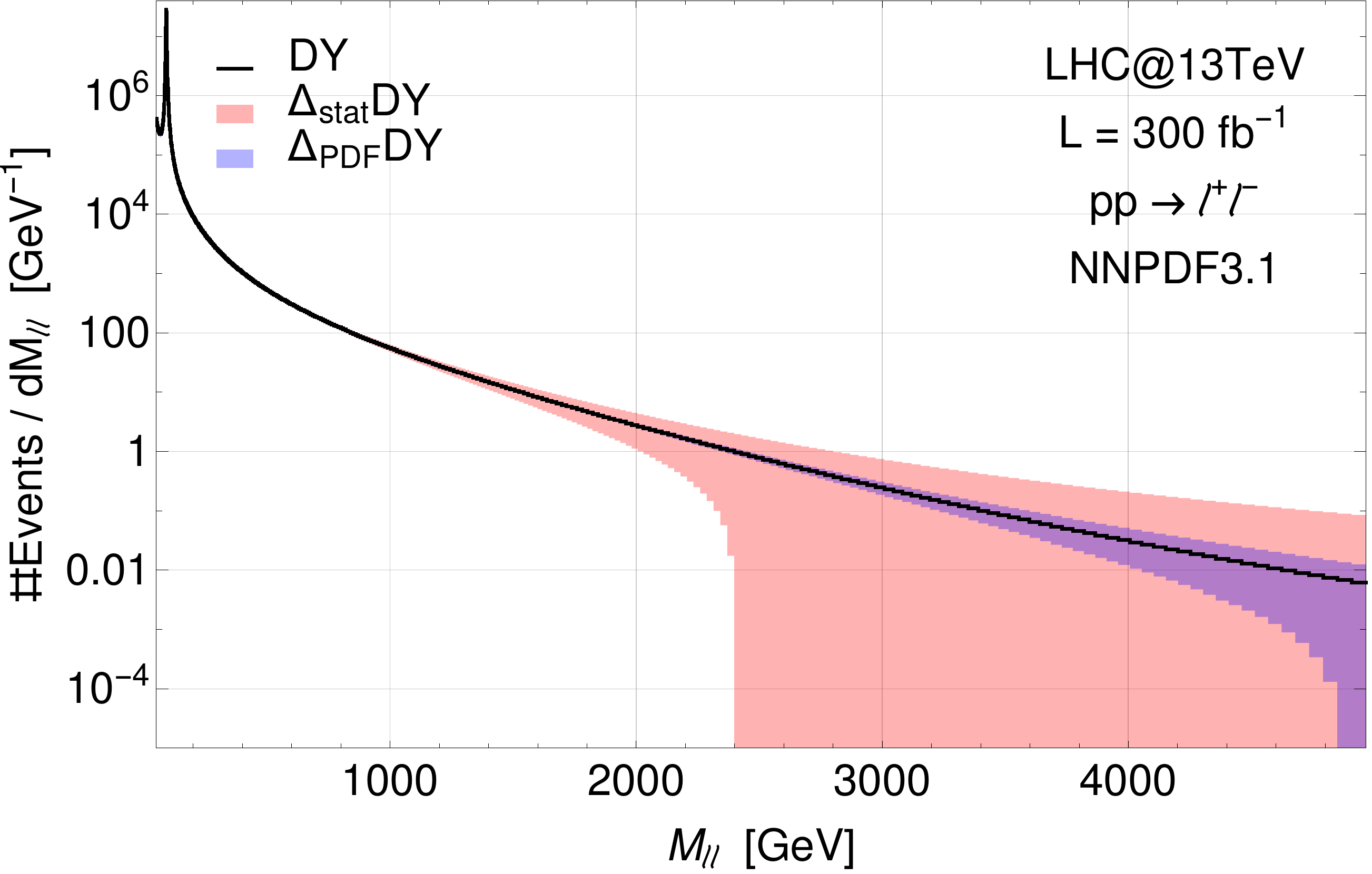}{(b)}
\includegraphics[width=0.47\textwidth]{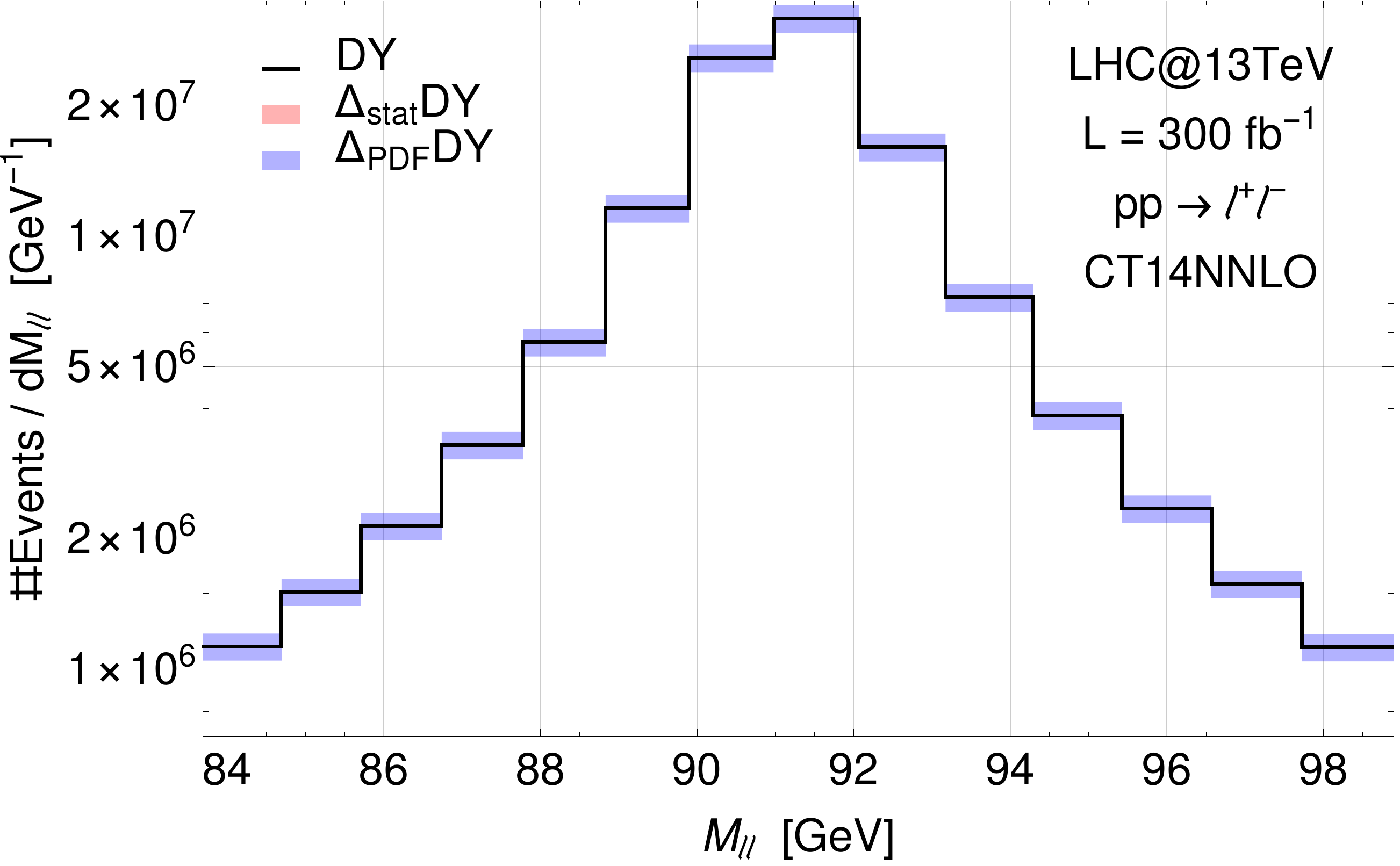}{(c)}
\includegraphics[width=0.47\textwidth]{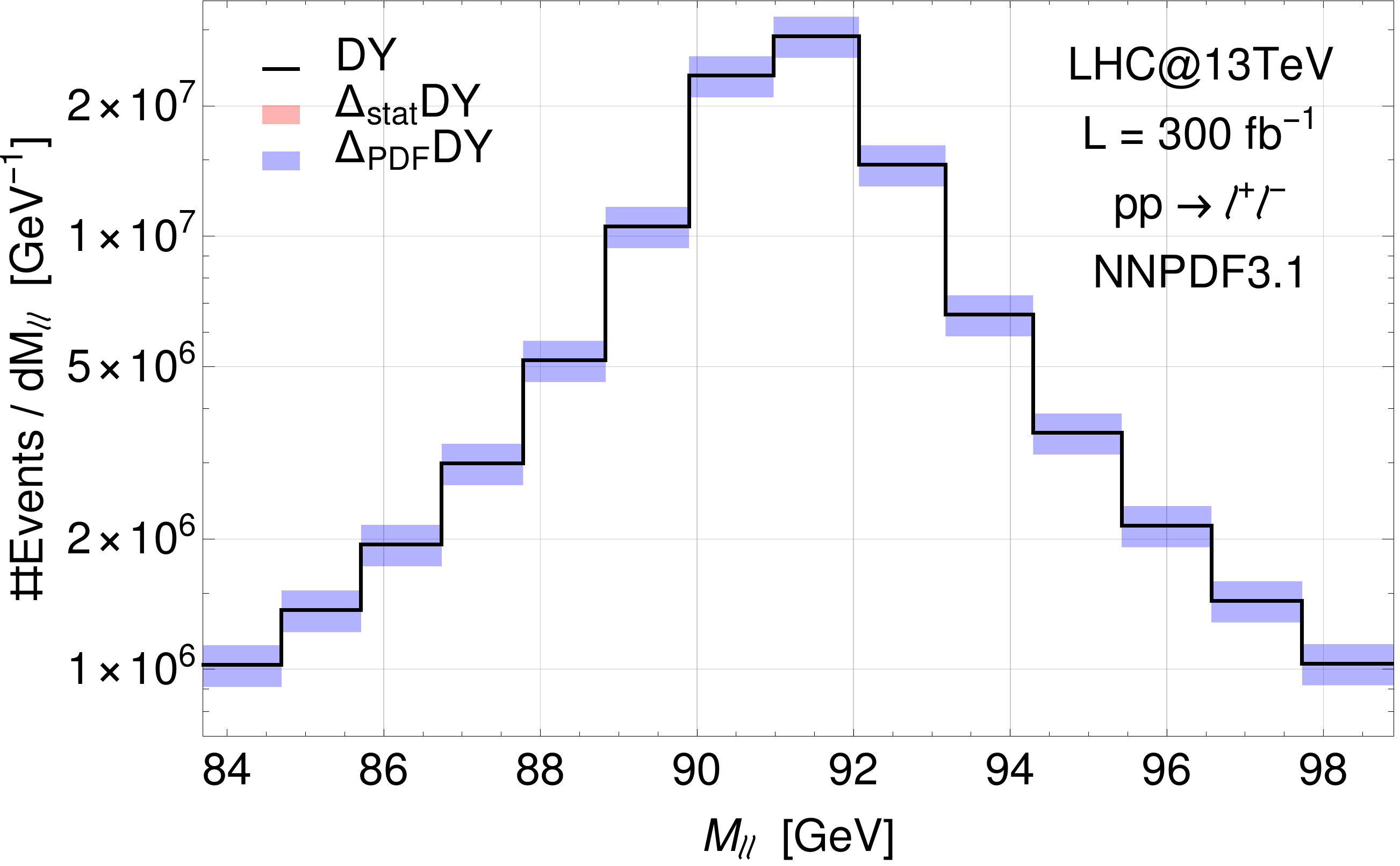}{(d)}
\caption{(a) Expected number of events in the di-lepton invariant mass region and their statistical and PDF error obtained with CT14NNLO. (b) Same as (a) obtained with NNPDF3.1.
(c) Zoom of (a) around the $Z$-boson peak. (d) Zoom of (b) around the $Z$-boson peak.}
\label{fig:XS_DY}
\end{center}
\end{figure}

The plan of our work is as follows. In Sec.~\ref{sec:calculation} we describe our calculation of the differential cross section and FB asymmetry in the NC DY channel.
In Sec.~\ref{sec:RunII} we compare statistical and PDF errors on these two observables. 
In Sec.~\ref{sec:PDFsets} we discuss the cut on the rapidity of the di-lepton system in measuring $A_{\rm FB}$ and its role in discriminating the different PDF sets.
We conclude in Sec.~\ref{sec:summa}.

\section{Differential Cross Section and FB asymmetry}
\label{sec:calculation}

In this section we focus on the NC DY process at the CERN LHC Run-II
\begin{center}
$pp\rightarrow\gamma^*, Z\rightarrow e^+e^-$
\end{center}
and we briefly describe how the differential cross section and FB asymmetry are computed.
We consider the LHC with centre-of-mass energy $\sqrt{s}$ = 13 TeV for the present and project values of the integrated luminosity. 

We calculate the number of expected events for the di-lepton cross section by including QCD Next-to-Next-to-Leading Order (NNLO) corrections~\cite{Hamberg:1990np, Harlander:2002wh}
and experimental acceptance and efficiency factors~\cite{Khachatryan:2014fba}.
We neglect the additional NLO EW corrections as they give rise to a K-factor estimated to be less than 1.035 in the di-lepton mass range considered (see Ref.~\cite{Aaboud:2017ffb}).
The residual scale uncertainty, after the inclusion of the available higher order corrections, is of the order of a few percent~\cite{Aaboud:2017ffb} and will not be considered here.
We perform calculations for two sample PDF sets, CT14NNLO~\cite{Dulat:2015mca} and NNPDF3.1~\cite{Ball:2017nwa}.
We compute the PDF uncertainties according to the methods described in detail in Refs.~\cite{Accomando:2015cfa,Accomando:2016tah,Accomando:2016ehi} and references therein. 
 
In Fig.~\ref{fig:XS_DY} we show the di-lepton cross section obtained with the two PDF sets considered.
We include both the expected statistical error and the PDF uncertainty for the standard luminosity $L =$ 300 fb$^{-1}$.
For low and intermediate invariant masses, $M_{\ell\ell}\le$ 700 GeV, the PDF uncertainty dominates in magnitude over the statistical error.
This is displayed more clearly by the (c)-(d) plots around the $Z$-boson peak.
For higher invariant masses, the statistics gets so poor at fixed luminosity that the statistical error becomes dominant.
Note that this is so even though the PDF uncertainty increases with the mass scale, as one enters the large-$x$ region. 

\begin{figure}[t]
\begin{center}
\includegraphics[width=0.47\textwidth]{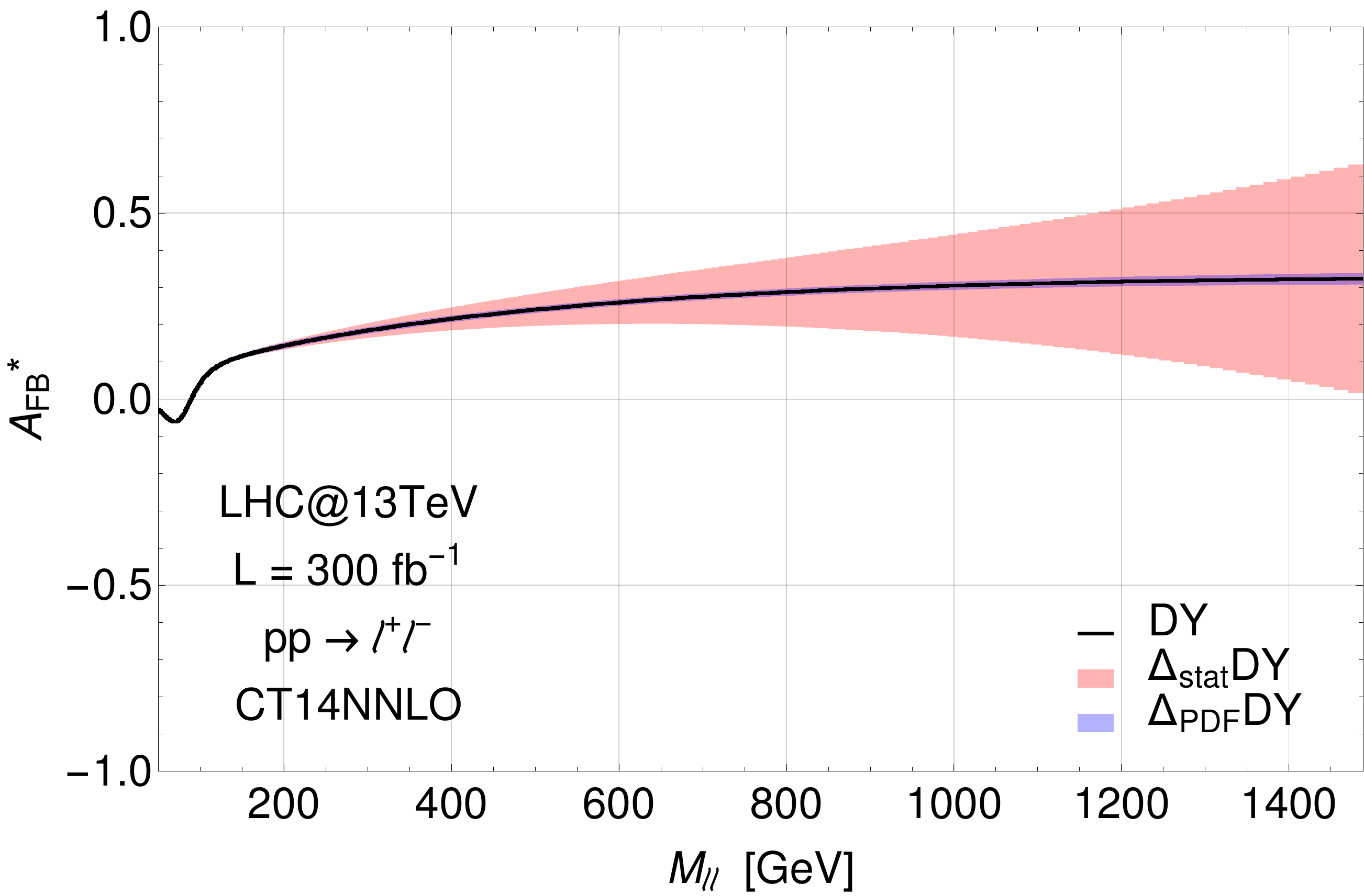}{(a)}
\includegraphics[width=0.47\textwidth]{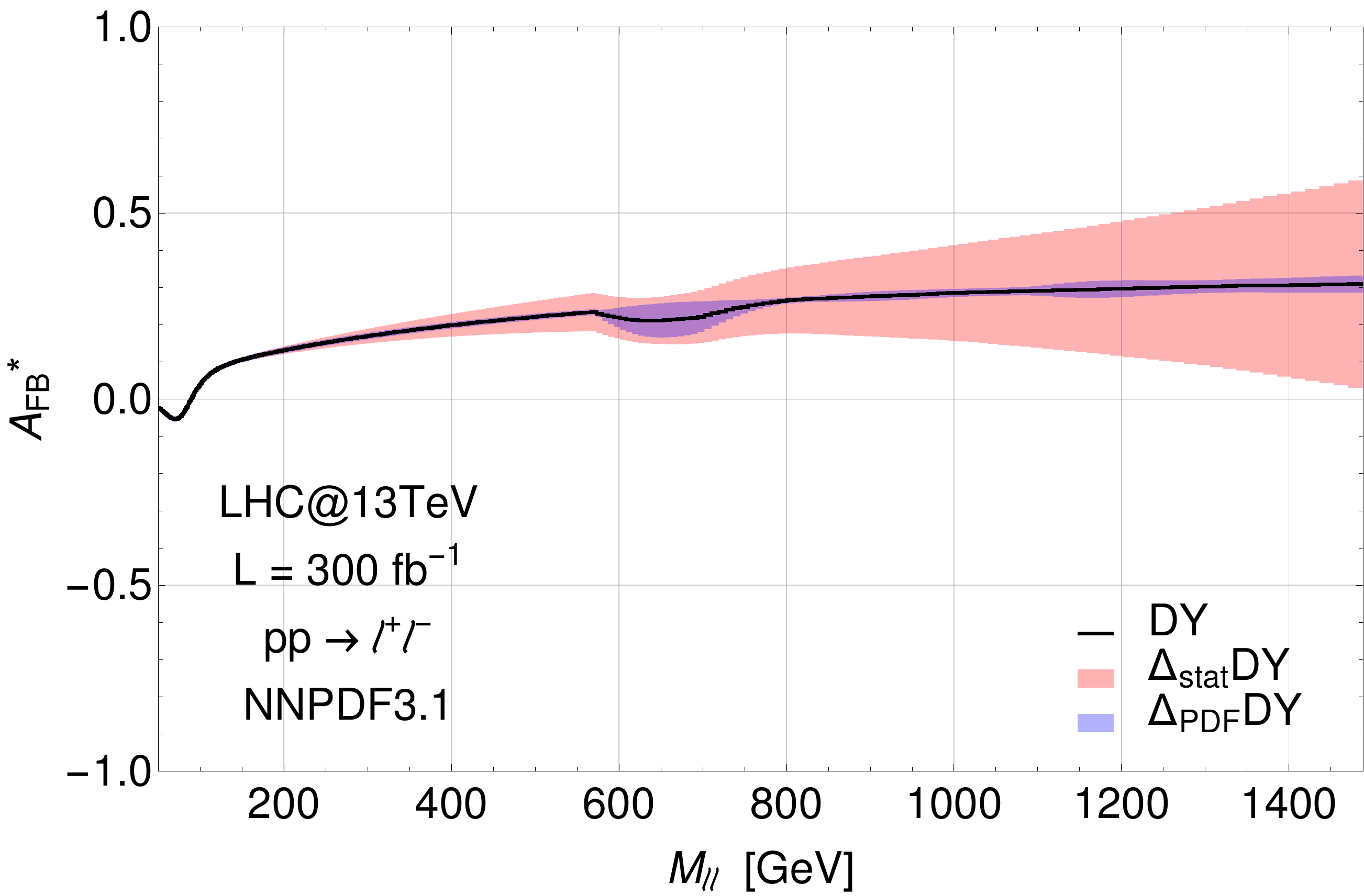}{(b)}
\includegraphics[width=0.47\textwidth]{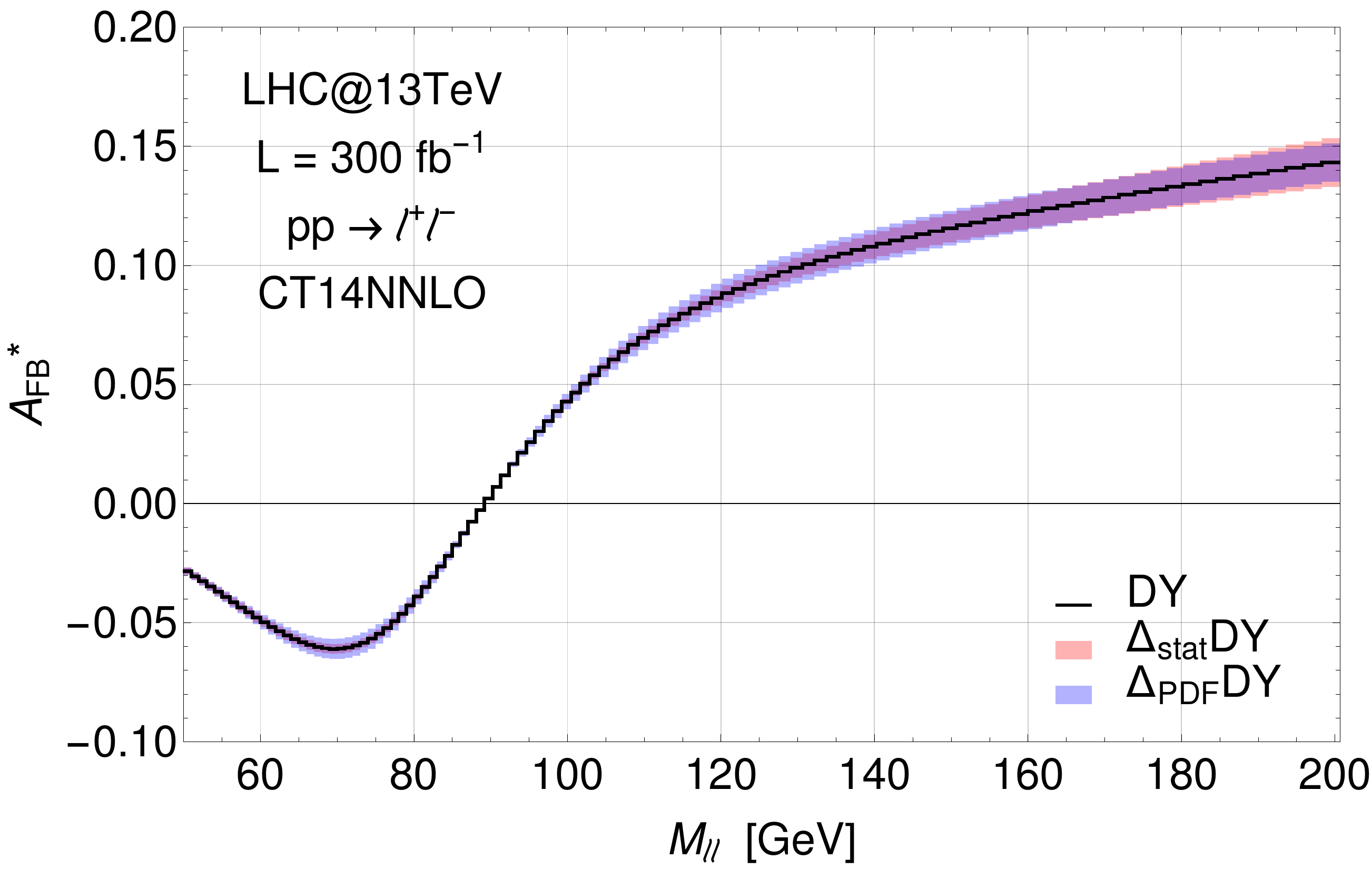}{(c)}
\includegraphics[width=0.47\textwidth]{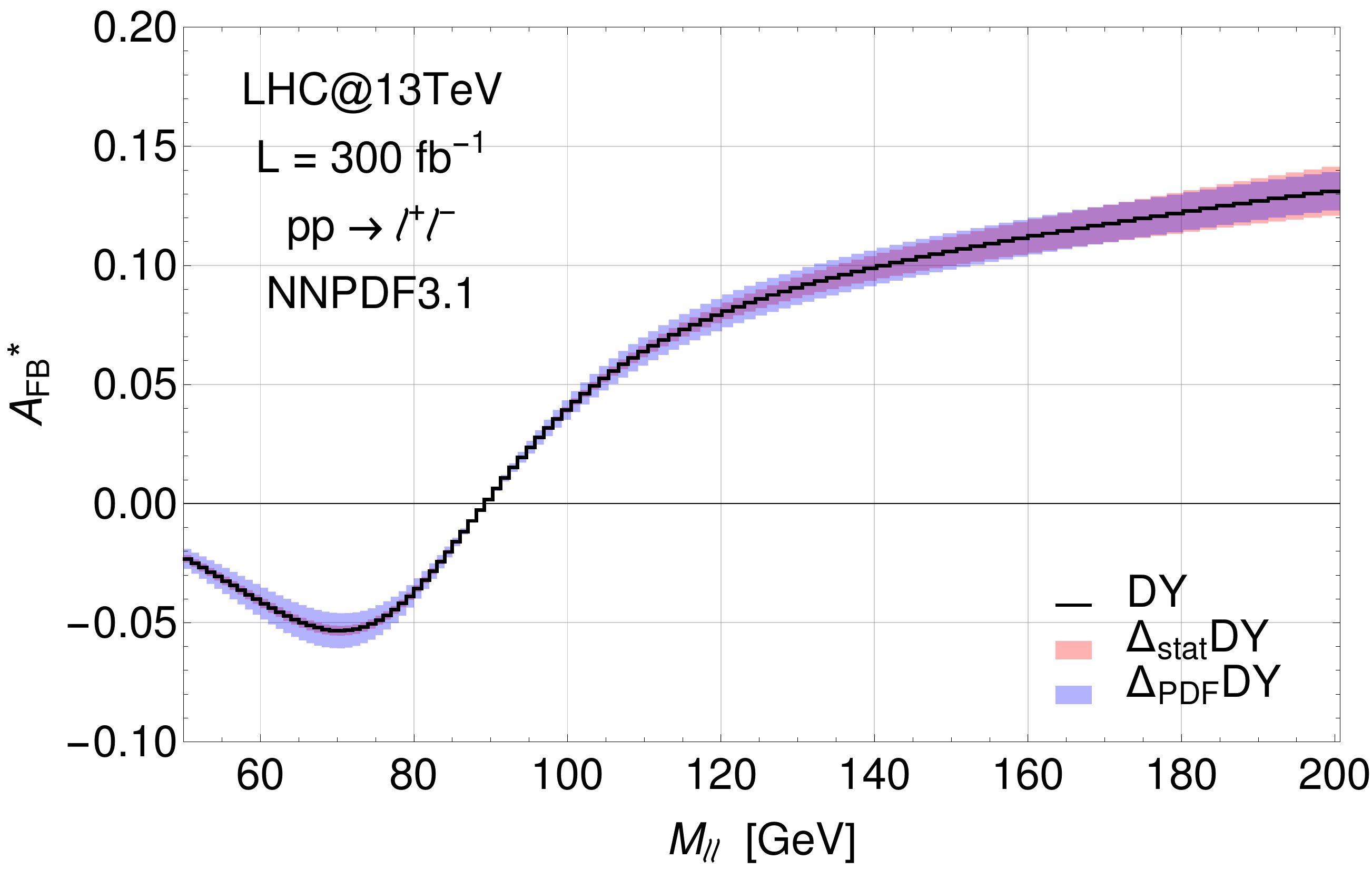}{(d)}
\caption{(a) $A_{\rm FB}^*$ distribution in the di-lepton invariant mass including the statistical and PDF errors obtained with CT14NNLO. (b) Same as (a) for the NNPDF3.1 PDF set.
(c) Zoom of (a) around the $Z$-boson peak. (d) Zoom of (b) around the $Z$-boson peak.}
\label{fig:AFB_DY}
\end{center}
\end{figure}

An analogous result is obtained for the FB asymmetry.
As the proton-proton collider does not allow one to directly access the direction of the incoming (anti)quark, we employ the method given in Refs.~\cite{Accomando:2015cfa,Dittmar:1996my,Rizzo:2009pu}
of using the boost direction of the di-lepton system to extract information on the (anti)quark axis.
As a measure of this boost, we define the di-lepton rapidity 
\begin{equation}
Y_{\ell\ell} = {1\over 2}{\rm ln}\Big [{{E+P_z}\over{E-P_z}}\Big ],
\end{equation}
where $E$ and $P_z$ are respectively the energy and the longitudinal momentum of the two final-state electrons.
We refer the reader to Ref.~\cite{Accomando:2015cfa} for details on the algorithm via which the FB asymmetry, $A^*_{\rm FB}$, is reconstructed and on the method for evaluating its PDF uncertainty.
In Fig.~\ref{fig:AFB_DY} we plot the (reconstructed) FB asymmetry as a function of the di-lepton mass scale.
We see that at fixed luminosity the PDF uncertainty dominates over the statistical uncertainty for $M_{\ell\ell}\le$ 170 GeV.
For higher masses, the statistical error takes over.
These results do not depend on the particular PDF set used, as we will confirm later.

\section{Prospects for PDF and statistical uncertainties at the LHC Run-II}
\label{sec:RunII}

\begin{figure}[t]
\begin{center}
\includegraphics[width=0.47\textwidth]{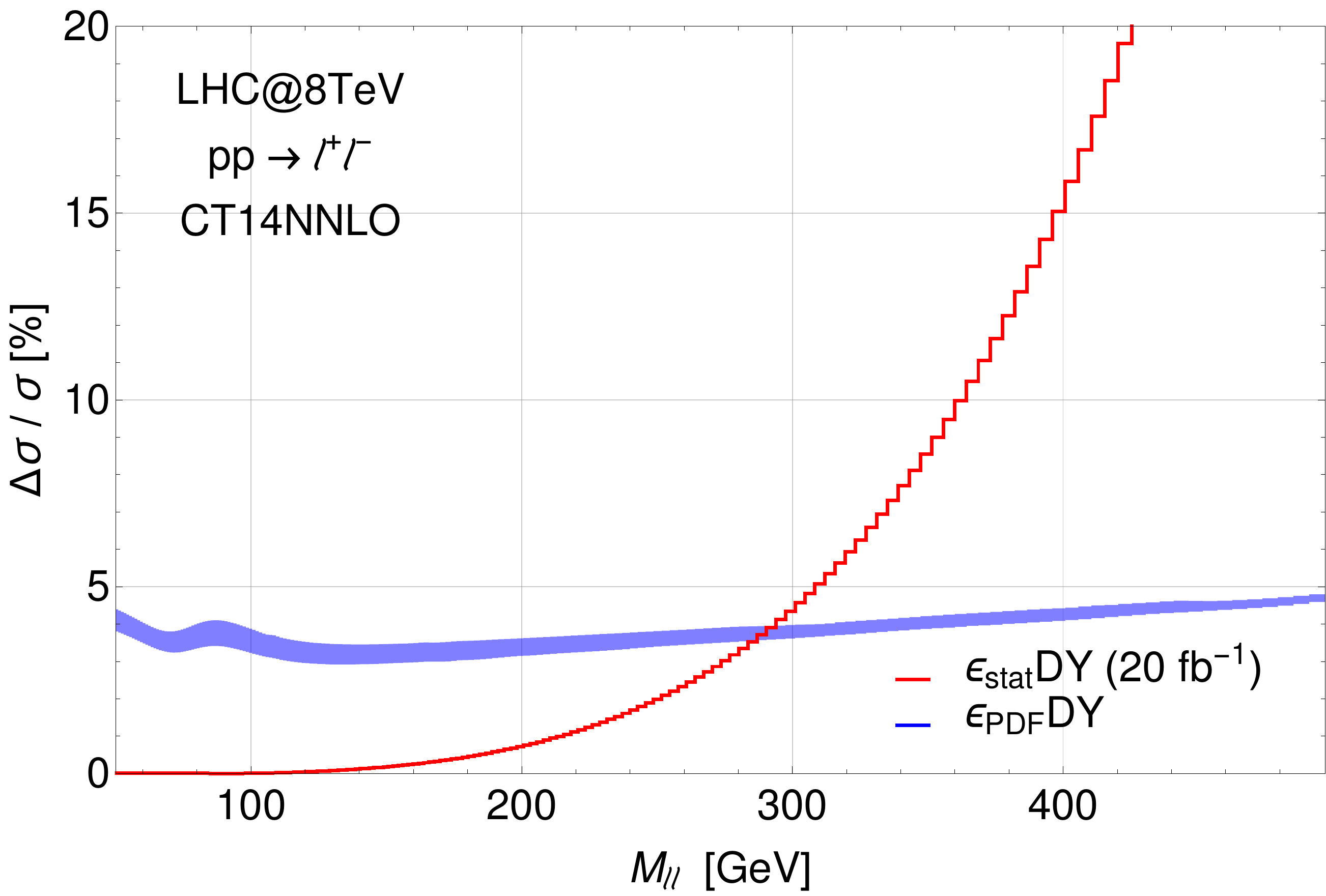}{(a)}
\includegraphics[width=0.47\textwidth]{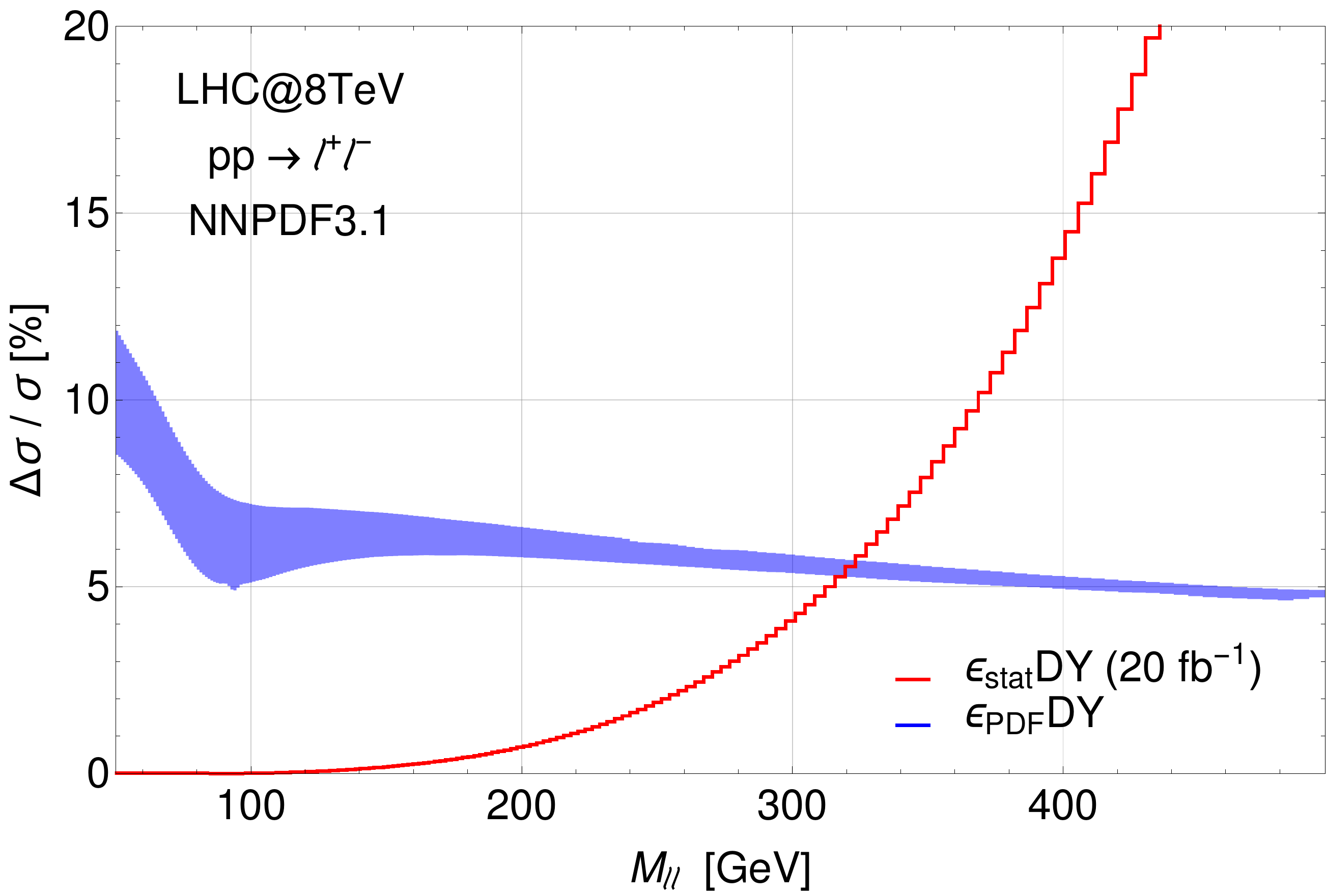}{(b)}
\includegraphics[width=0.47\textwidth]{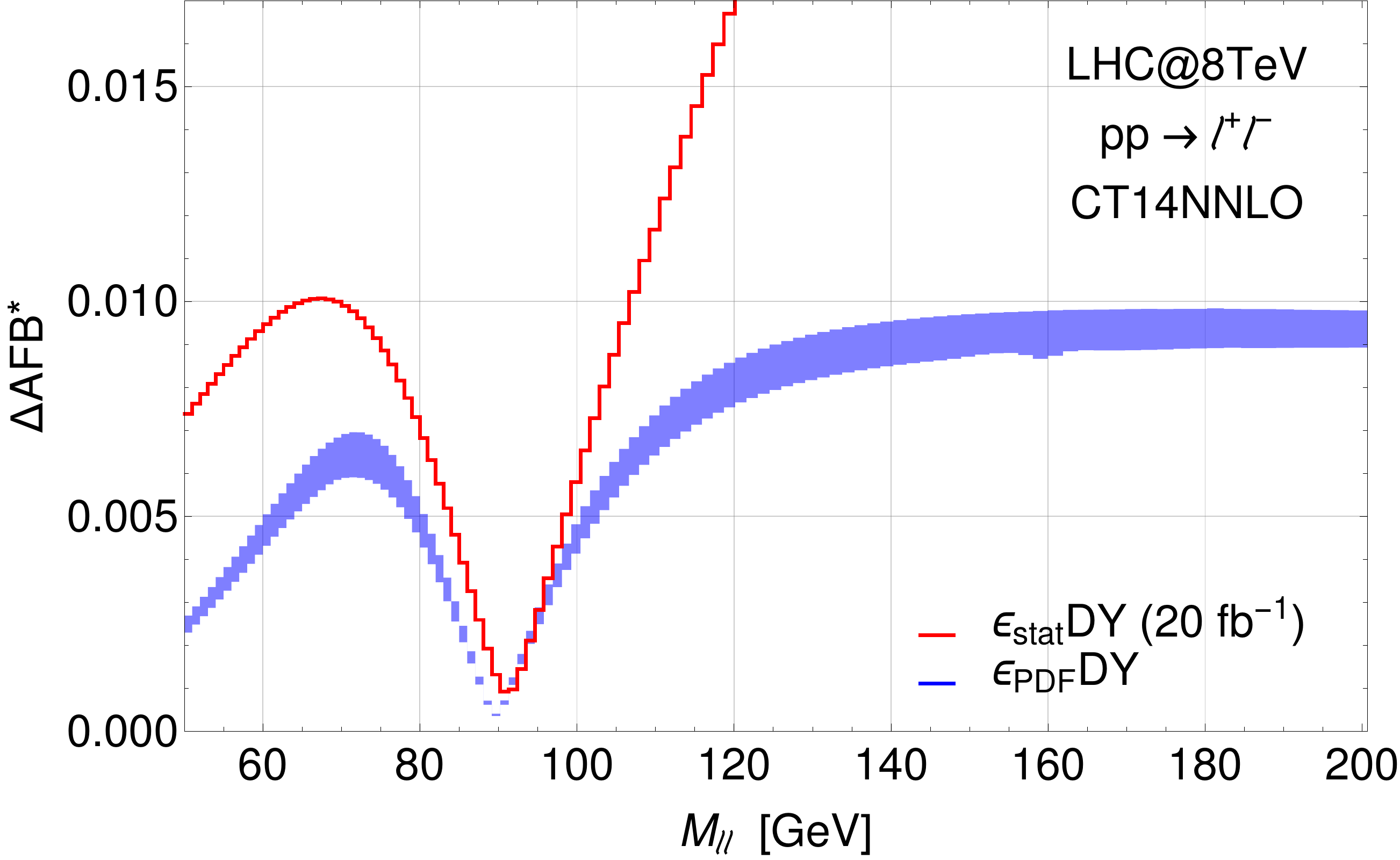}{(c)}
\includegraphics[width=0.47\textwidth]{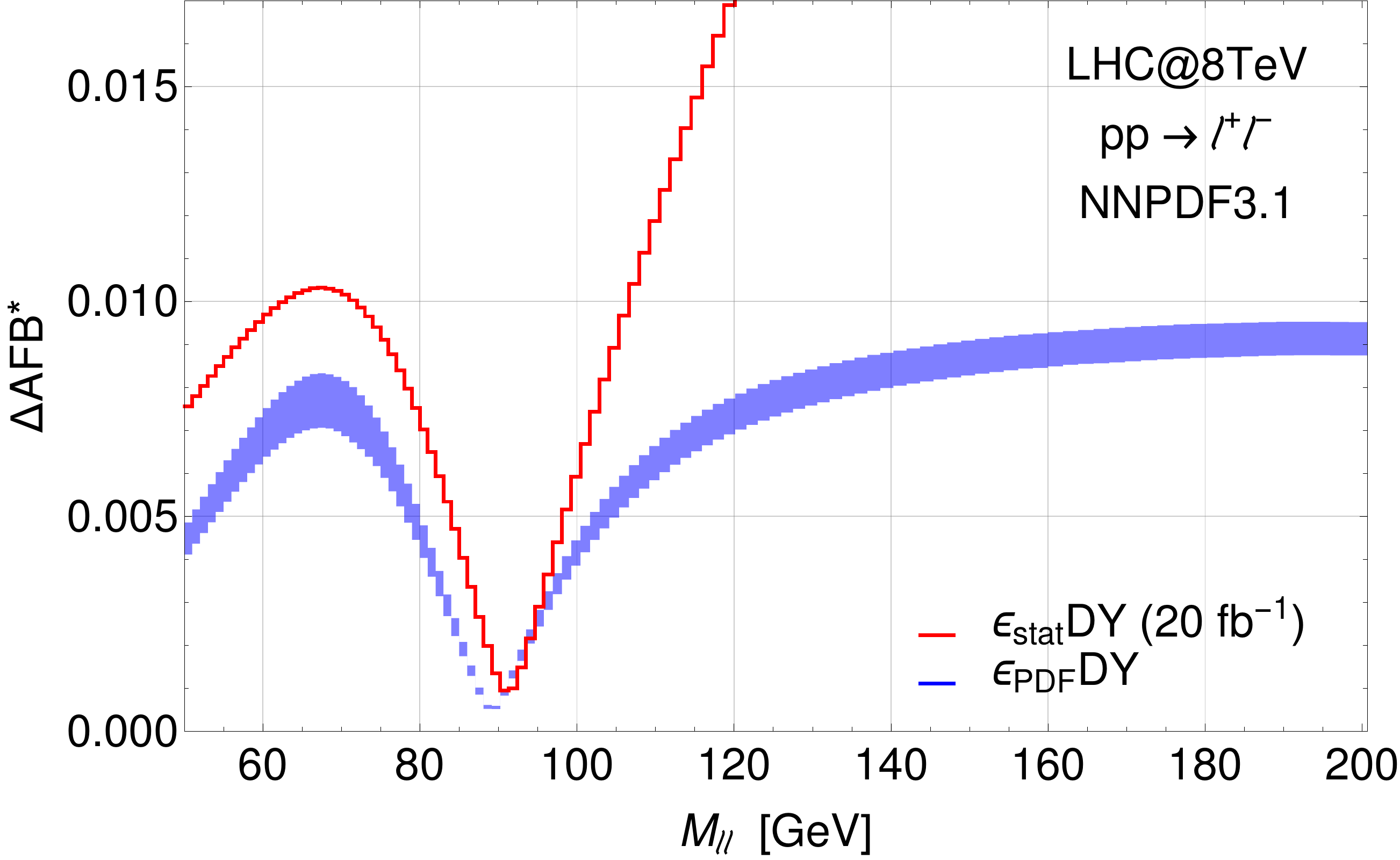}{(d)}
\caption{(a) Relative errors on the differential cross section at the LHC Run-I with $\sqrt{s}$ = 8 TeV and $L =$ 20 fb$^{-1}$, computed with the CT14NNLO PDF set.
The red curve represents the statistical error.
The blue band has been obtained evaluating the PDF error fixing the factorization/renormalization scale in the interval: $0.5 M_{\ell\ell} < Q < 2 M_{\ell\ell}$.
(b) Same as (a) for the NNPDF3.1 PDF set.
(c) Absolute errors on the reconstructed FB asymmetry at the LHC Run-I, computed with the CT14NNLO PDF set.
The red curve represents the statistical error. The blue band refers to the PDF uncertainty. (d) Same as (c) for the NNPDF3.1 PDF set.}
\label{fig:8TeV}
\end{center}
\end{figure}

In this section we examine the evolution of the PDF and statistical errors on the di-lepton cross section and FB asymmetry from Run-I to Run-II of the LHC and propose the $A^*_{\rm FB}$ observable
to improve determinations of (anti)quark PDFs. 
Although we illustrate this by presenting explicit numerical results for the two representative PDF sets CT14NNLO and NNPDF3.1, our point is general and applies to other PDF sets as well.

In Fig.~\ref{fig:8TeV} we compare the statistical and PDF relative error on the di-lepton differential cross section (plots (a) and (b))
and the corresponding absolute errors on the reconstructed FB asymmetry (plots (c) and (d)) at the LHC Run-I with energy $\sqrt{s}$ = 8 TeV and integrated luminosity $L =$ 20 fb$^{-1}$.
We use the CT14NNLO PDF set in plots (a) and (c) plus the NNPDF3.1 PDF set in plots (b) and (d).
The extremes of the PDF uncertainty error band represent the PDF error calculated accordingly to the prescription adopted within each PDF set (eigenvectors for CT14NNLO and replicas for NNPDF3.1)
and fixing the factorization/renormalization scale equal to $\mu_F = \mu_R = 0.5 M_{\ell\ell}$ and $\mu_F = \mu_R = 2 M_{\ell\ell}$ respectively.
The same applies in all the plots that will be presented in the following.
The aim is to show that our conclusions following the comparison of statistical and PDF uncertainties are robust against the choice of this scale.

Figure~\ref{fig:8TeV} shows that the statistical accuracy is better than the PDF uncertainty in the low mass part of the $M_{\ell\ell}$ spectrum, especially around the $Z$-boson peak.
The inclusion of the LHC data up to $M_{\ell\ell}\le$ 280 - 320 GeV in the PDF fits could then improve, as expected, the extraction of the (anti)quark PDFs.
Conversely, for $A_{\rm FB}^*$ the two sources of uncertainty are rather similar in magnitude (see plots (b) and (c)).
Therefore, no gain in the (anti)quark PDF precision should be expected from the reconstruction of this observable with the data collected at $\sqrt{s}$ = 8 TeV.

\begin{figure}[t]
\begin{center}
\includegraphics[width=0.47\textwidth]{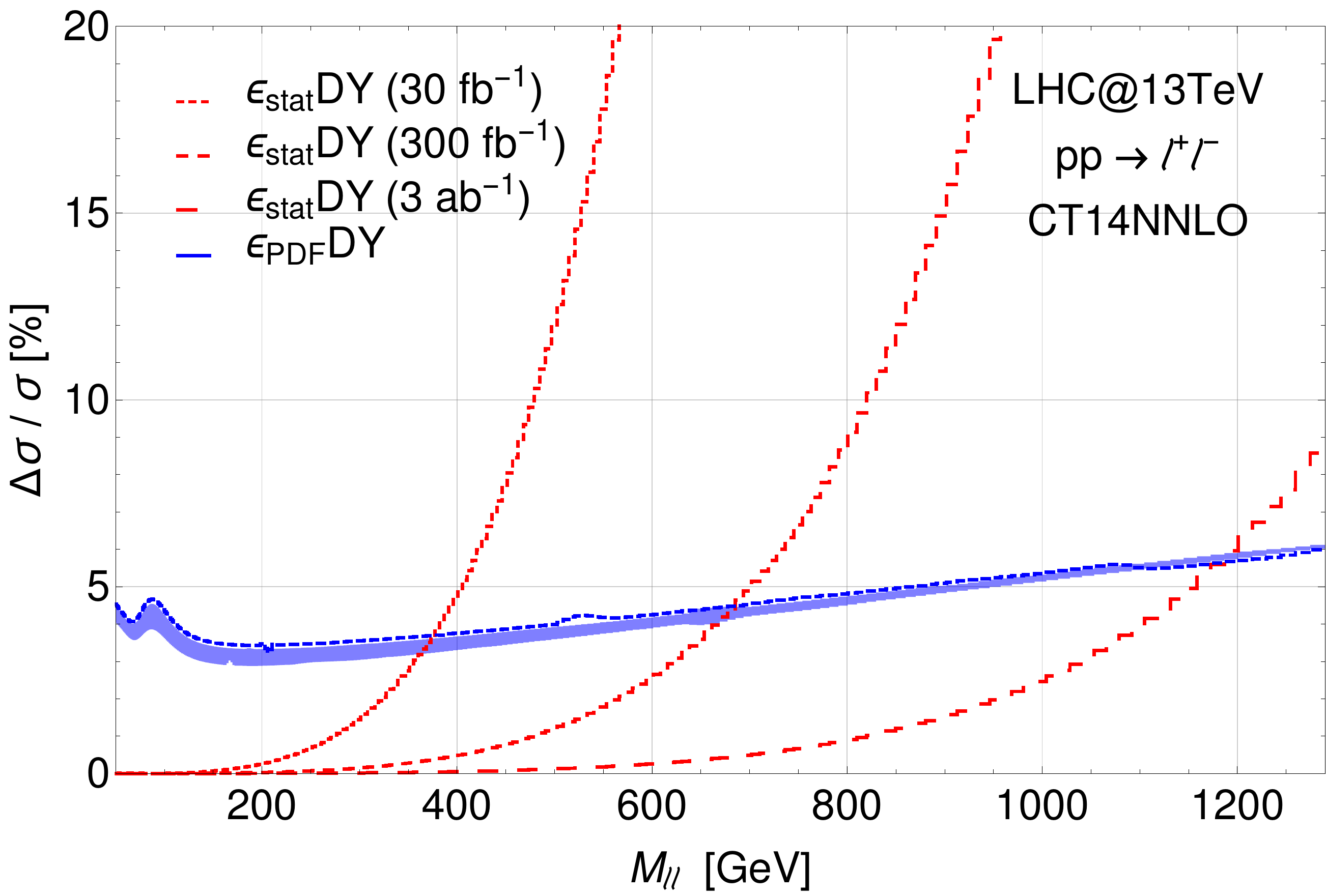}{(a)}
\includegraphics[width=0.47\textwidth]{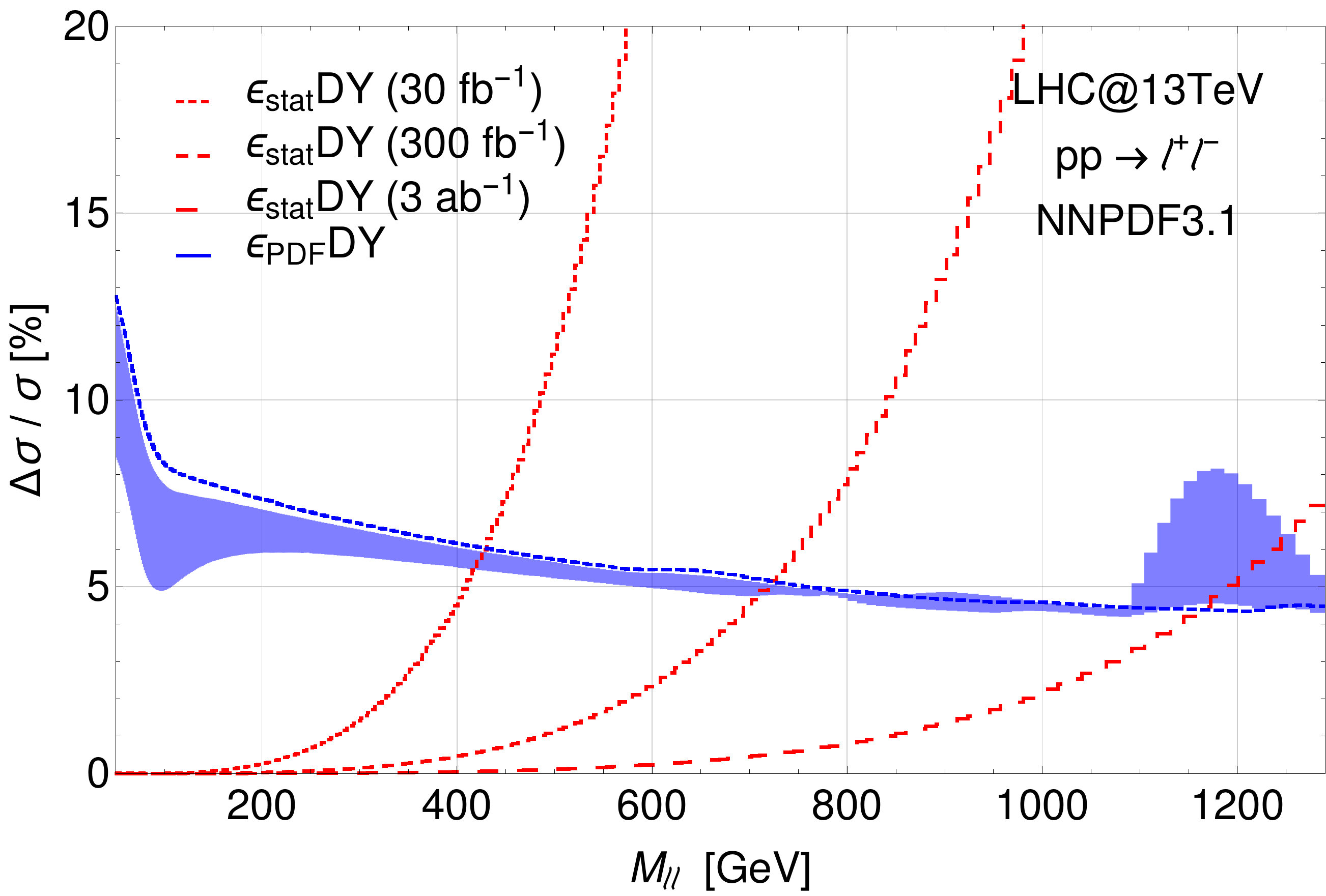}{(b)}
\includegraphics[width=0.47\textwidth]{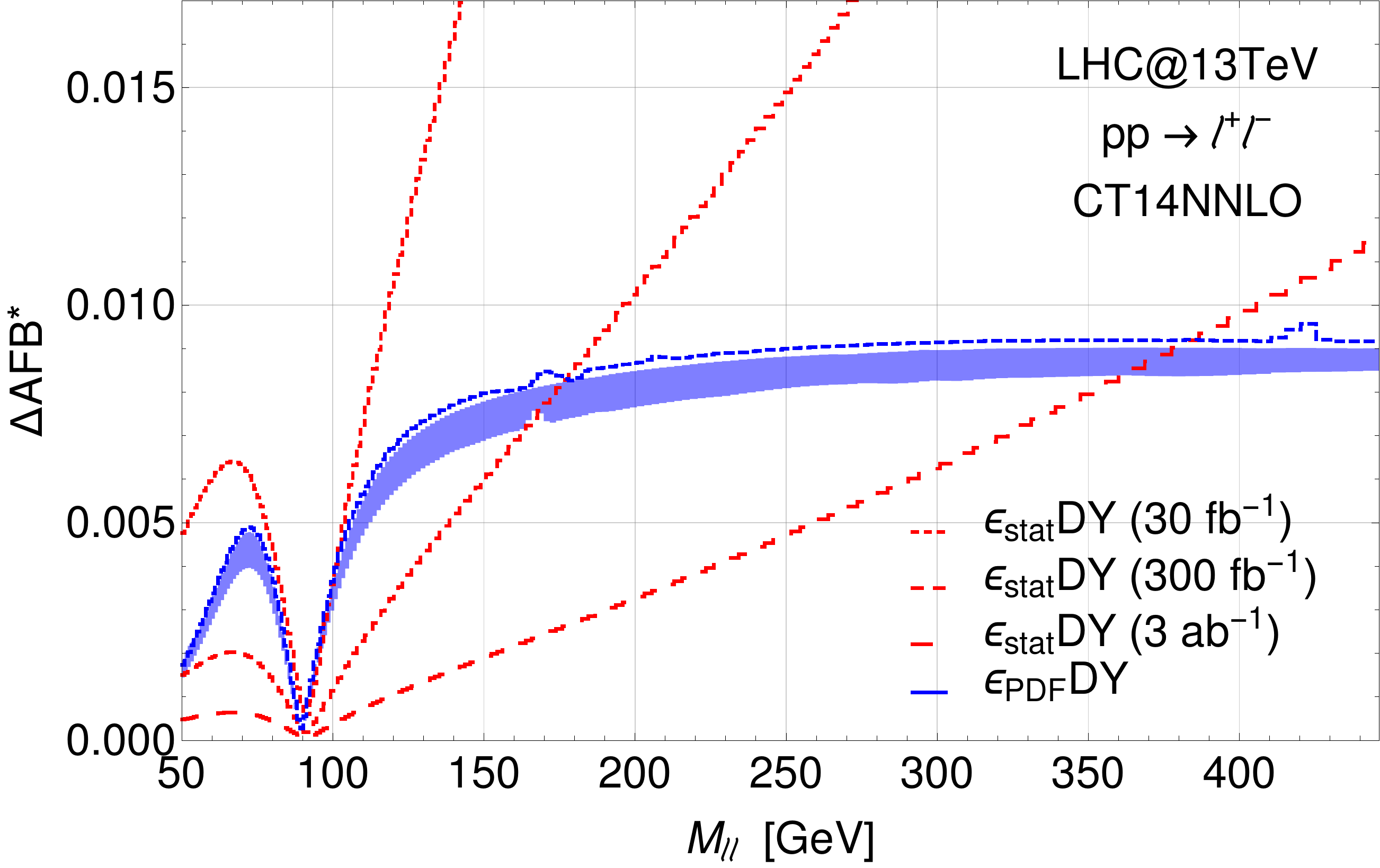}{(c)}
\includegraphics[width=0.47\textwidth]{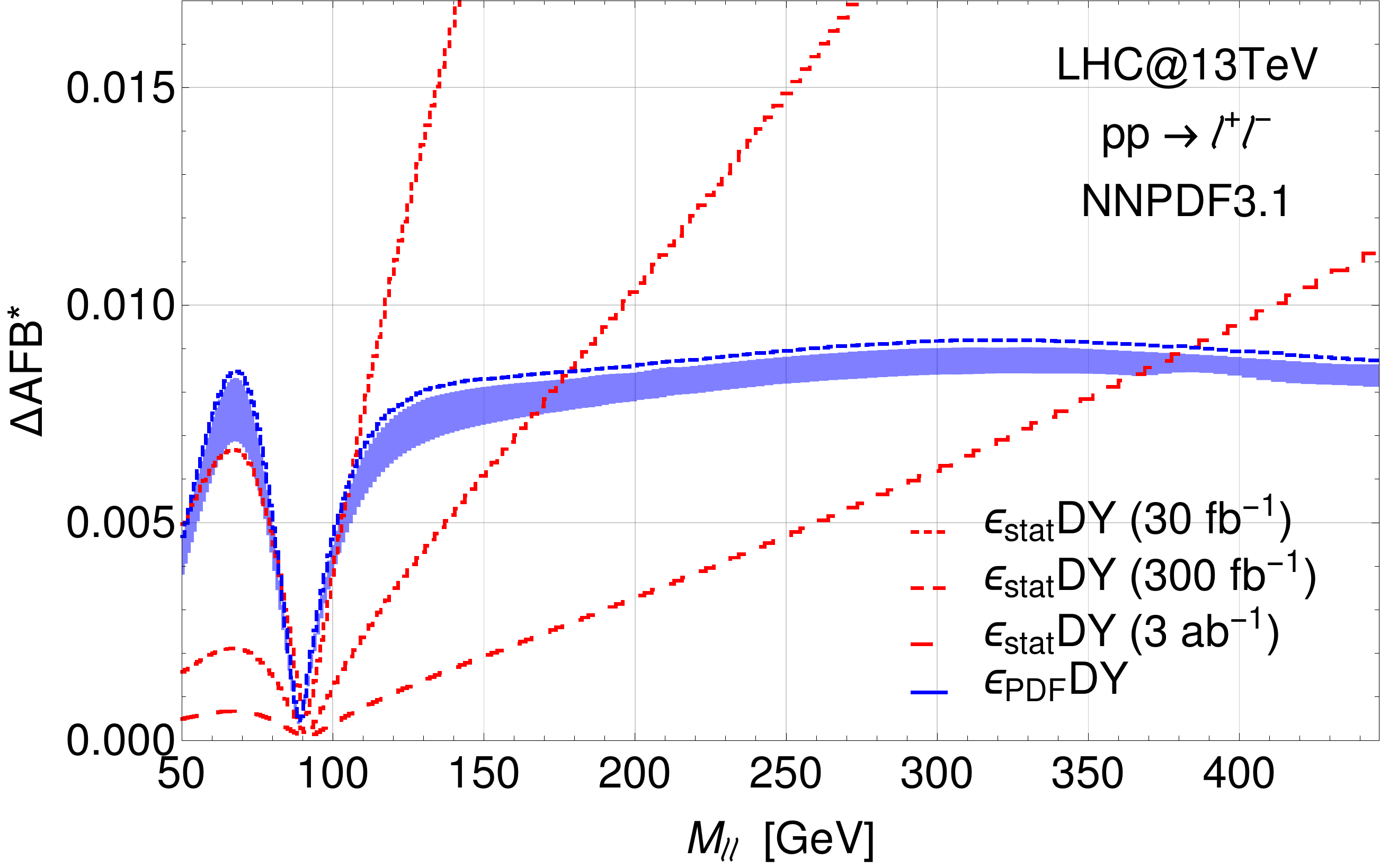}{(d)}
\caption{(a) Relative size of statistical error and PDF uncertainty on the differential cross section obtained with the CT14NNLO set at the 13 TeV LHC.
The statistical error is displayed for three different values of the luminosity (see legend).
The PDF error band has been obtained evaluating the PDF error fixing the factorization/renormalization scale in the interval: $0.5 M_{\ell\ell} < Q < 2 M_{\ell\ell}$,
while the dashed blue lines represent the PDF uncertainty obtained fixing $\mu_F = \mu_R = p_T$, with $p_T$ the transverse momentum of either lepton in the final state.
(b) Same as plot (a) for the NNPDF3.1 PDF set.
(c) Absolute value of statistical and PDF uncertainties on the reconstructed $A_{\rm FB}^*$ distribution obtained with CT14NNLO.
The statistical error is obtained with the integrated luminosities of 30 fb$^{-1}$, 300 fb$^{-1}$ and 3 ab$^{-1}$. The PDF error bands and lines follow the conventions of plots (a) and (b).
(d) Same as (c) for the NNPDF3.1 PDF set.}
\label{fig:DY_HL}
\end{center}
\end{figure}

However, the situation changes at the present LHC Run-II. This is noticeable in Fig.~\ref{fig:DY_HL}.
Again the extremes of the PDF error bands have been obtained calculating the PDF error following each PDF set prescription
and fixing the factorization/renormalisation scale equal to $\mu_F = \mu_R = 0.5 M_{\ell\ell}$ and $\mu_F = \mu_R = 2 M_{\ell\ell}$ respectively.
The dashed blue lines represent the PDF error obtained fixing the factorization/renormalisation scale $\mu_F = \mu_R = p_T$,
where $p_T$ is the transverse momentum of either lepton in the final state.
In the upper plots ((a) and (b)) we show the relative size of the statistical and PDF uncertainty as a function of the invariant mass for three different values of the luminosity $L =$ 30 fb$^{-1}$, 300 fb$^{-1}$ and 3 ab$^{-1}$.
From the two upper plots, one can see that the statistical error is lower than the PDF error in the low to intermediate mass region.
For the projected luminosity $L =$ 3 ab$^{-1}$ at the envisaged High-Luminosity (HL) LHC \cite{Gianotti:2002xx}, the statistical error is lower than the PDF error for $M_{\ell\ell} <$ 1150 GeV in both the CT14NNLO and NNPDF3.1 cases.

The reconstructed FB asymmetry, $A_{\rm FB}^*$, experiences a partial cancellation of PDF uncertainties. In this sense we expect that the statistics would play here a major role.
As visible from plots (c) and (d) of Fig.~\ref{fig:DY_HL}, in the region of intermediate to large invariant masses the statistical uncertainty is dominant, and this happens at lower invariant masses as compared to the case of the 
differential cross section (see plots (a) and (b)).
However in the region around the $Z$-boson peak, we find a similar situation as in the differential cross section case.
Despite the partial cancellation of systematical errors, in the case of the $A_{\rm FB}^*$ observable too there exists a region in the di-lepton invariant mass in which the statistical errors are lower than the PDF uncertainties.
Assuming again the projected luminosity $L =$ 3 ab$^{-1}$ of the HL LHC, this is true for roughly $M_{\ell\ell} <$ 370 GeV in both the CT14NNLO and NNPDF3.1 frameworks.
The off-peak region of the di-lepton cross section where the (anti)quark PDF uncertainty could be constrained by including the $A_{\rm FB}^*$ data at Run-II and HL stages is smaller than the analogous region
where the differential cross section data could be used.
However, the major strength of the FB asymmetry is its robustness against systematical errors.
From this point of view, the measured $A_{\rm FB}^*$ could be highly competitive with the (default) single-differential cross section and even with the more recent triple differential cross section from which it can be obtained. For the $A_{\rm FB}^*$ observable, both the uncorrelated and correlated components of the uncertainty are determined and then summed up in quadrature. The result is that the global error is significantly reduced when compared to the triple differential cross section. The total uncertainty is dominated by the data statistical one everywhere ~\cite{Aaboud:2017ffb}.
A further important point is that, typically, the PDF error is sizeably reduced in the reconstructed asymmetry. This means that this observable could a priori be in a better position to display tensions between different PDF fits performed with different sets of data. Also, it could show discrepancies coming from different parametrizations within the same PDF set or between two (or more) distinct sets, as we discuss in the next section.

\section{Distinguishing between PDF sets}
\label{sec:PDFsets}

\begin{figure}[t]
\begin{center}
\includegraphics[width=0.47\textwidth]{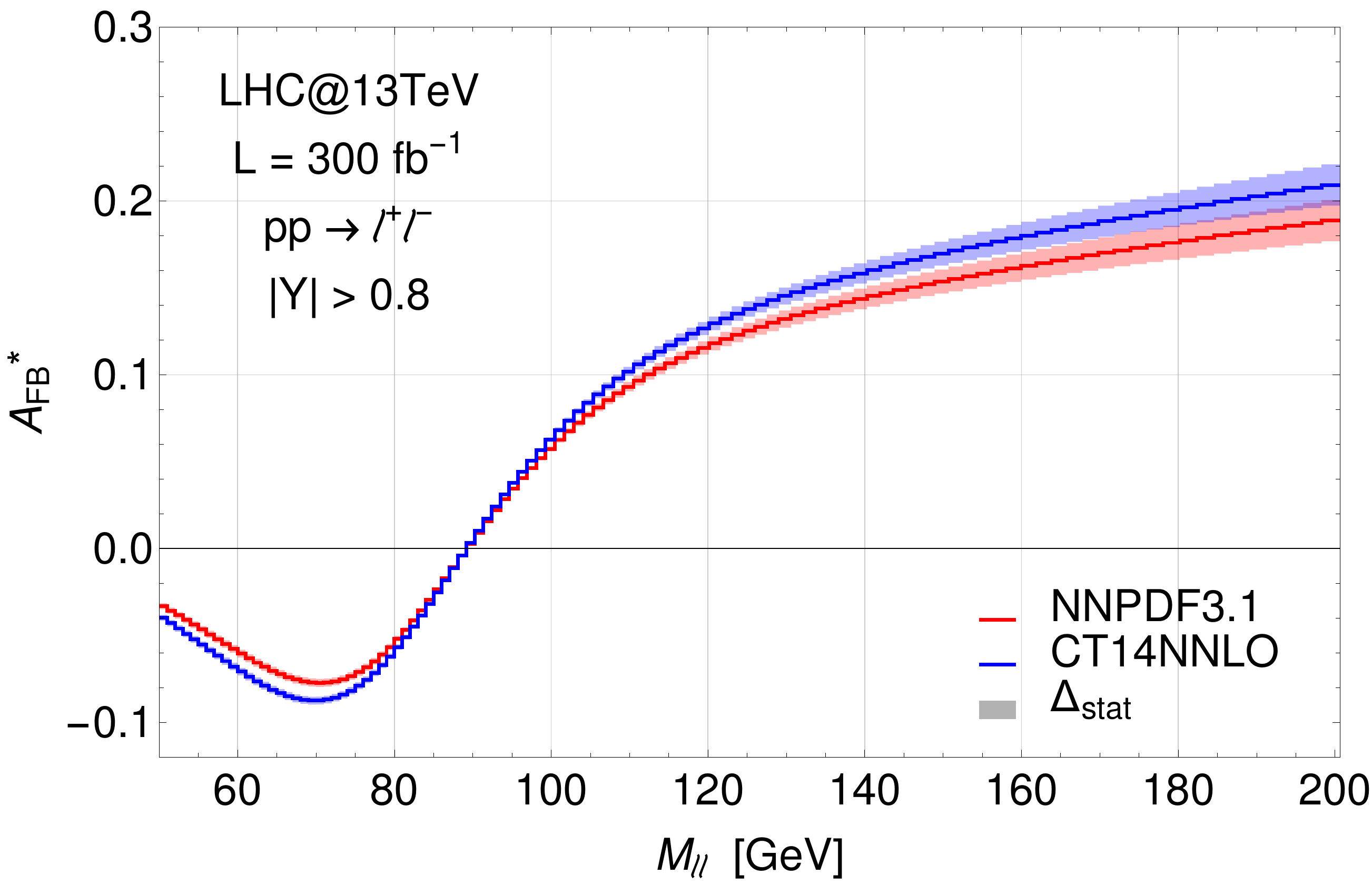}{(a)}
\includegraphics[width=0.47\textwidth]{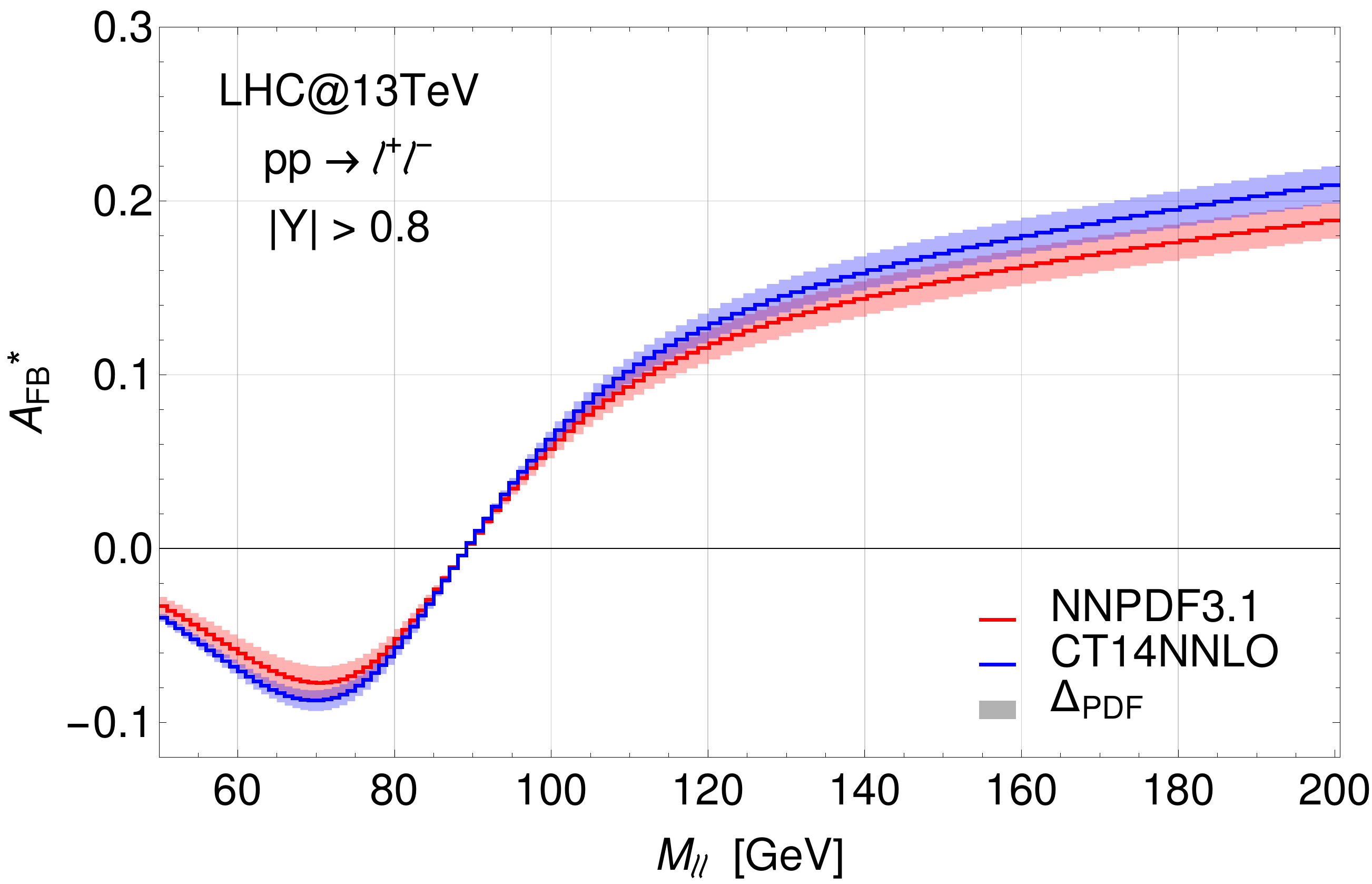}{(b)}
\includegraphics[width=0.47\textwidth]{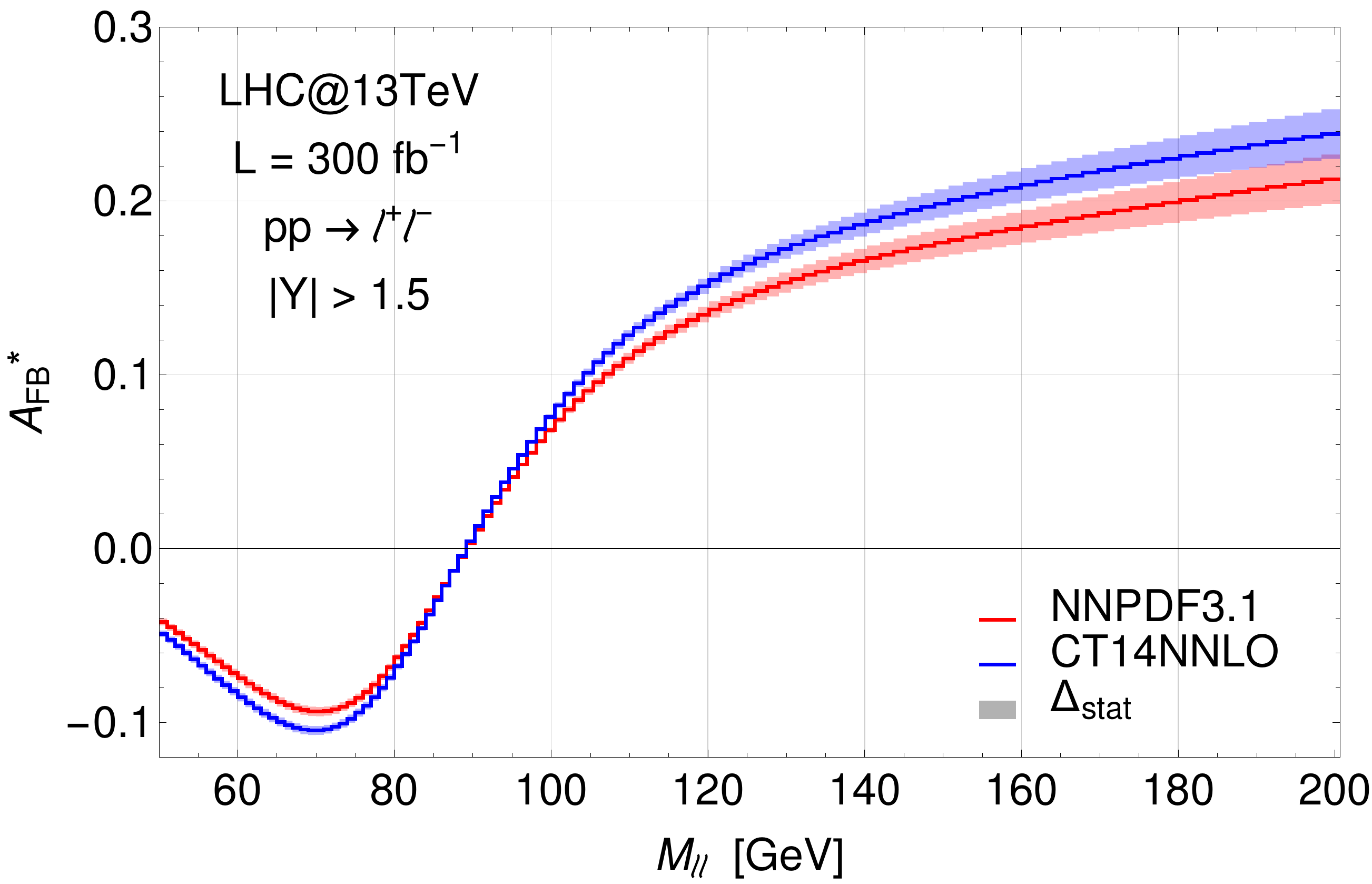}{(c)}
\includegraphics[width=0.47\textwidth]{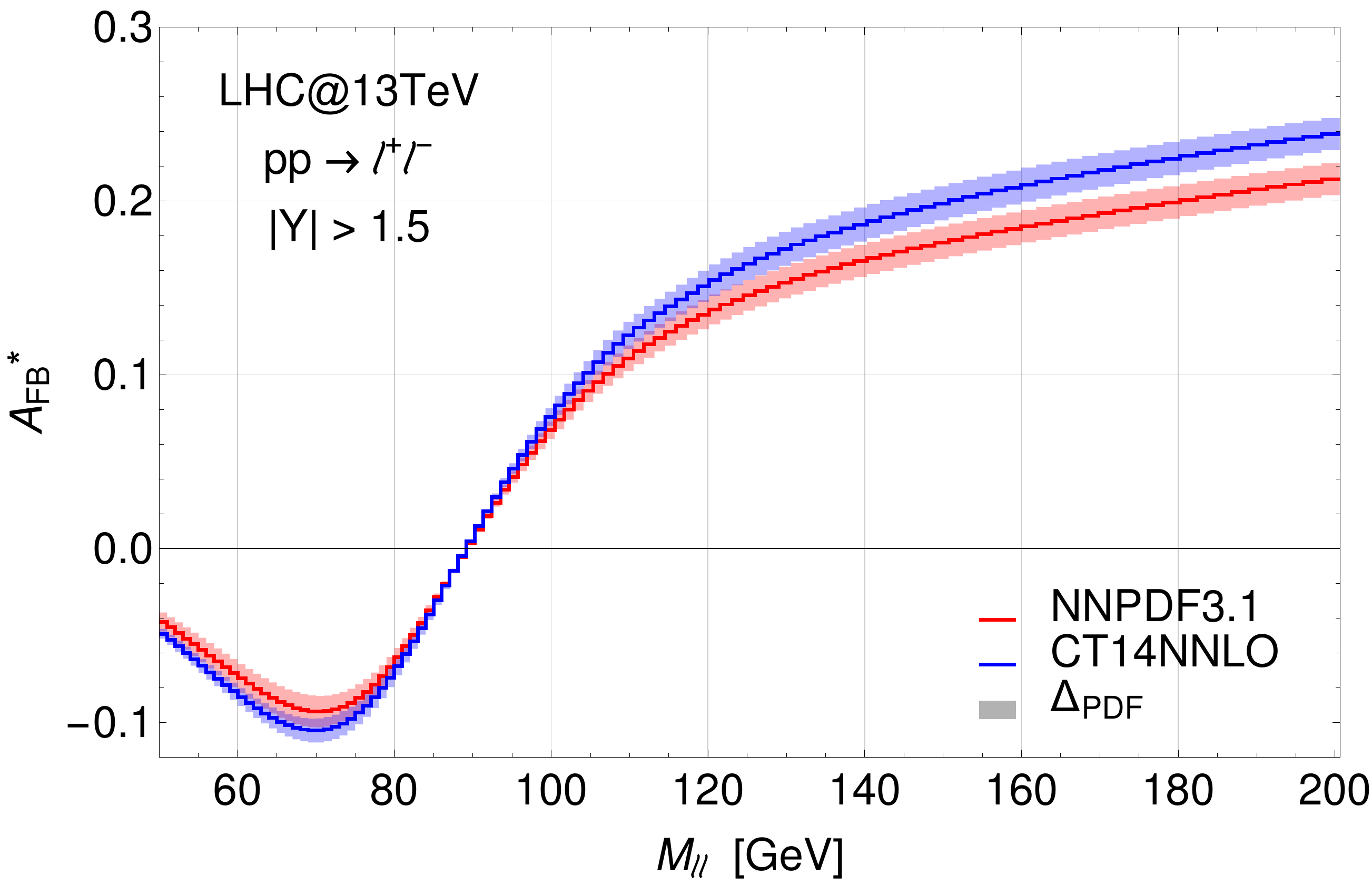}{(d)}
\caption{(a) $A_{\rm FB}^*$ distribution in the invariant mass region around the Z peak at the LHC Run-II with $L =$ 300 fb$^{-1}$, computed with the CT14NNLO and the NNPDF3.1 PDF sets.
The di-lepton rapidity cut $|Y_{\ell\ell}| > 0.8$ is imposed. The uncertainty band represents the statistical error.
(b) Same as (a) with the PDF error band. (c) Same as (a) for the di-lepton rapidity cut $|Y_{\ell\ell}| > 1.5$. (d) Same as (b) for the di-lepton rapidity cut $|Y_{\ell\ell}| > 1.5$.}
\label{fig:AFB_Z_peak_y_cut}
\end{center}
\end{figure}

In this section we suggest a method to discriminate between the different parameterizations of the (anti)quark PDFs via the measurement of the reconstructed FB asymmetry.
We start by observing that applying a cut on the rapidity of the di-lepton system $Y_{\ell\ell}$ improves the efficiency in searching for the correct direction of the quark.
As a consequence, the reconstruction of the FB asymmetry is more accurate and one can recover a shape that more closely resembles the true $A_{\rm FB}$.
The change in the $A_{\rm FB}^*$ shape due to the rapidity cut depends on the partonic content of the proton and therefore it is different for each choice of PDF set.
In Fig.~\ref{fig:AFB_Z_peak_y_cut} we plot the $A_{\rm FB}^*$ distribution obtained with the CT14NNLO (plots (a) and (c)) and NNPDF3.1 (plots (b) and (d)) sets for two sample cuts on the 
di-lepton rapidity: $Y_{\ell\ell}$ = 0.8 and $Y_{\ell\ell}$ = 1.5.
In the plots on the left, we show the statistical error band. In the plots on the right, we show the PDF uncertainty band. 

We consider the value of luminosity at the end of Run-II, $L =$ 300 fb$^{-1}$.
We see from Fig.~\ref{fig:AFB_Z_peak_y_cut} that the discriminating potential of the reconstructed $A_{\rm FB}^*$ increases with increasing the cut on the di-lepton rapidity. In particular, the two bottom plots of Fig.~\ref{fig:AFB_Z_peak_y_cut} indicate that for $|Y_{\ell\ell}| > 1.5$ the two selected PDF sets could be distinguished beyond their own statistical and PDF errors.
This result is due to the fact that a stringent rapidity cut selects the region of large longitudinal momentum fractions $x$, where the PDFs are least known.
This application of the $A_{\rm FB}^*$ observable for discriminating between different PDF sets thus relies on the ability of combining the high statistics from a region of the spectrum still close enough
to the $Z$-peak with the sensitivity to the large-$x$ domain enabled by the stringent $Y_{\ell\ell}$ cut.

We note that selecting non-central di-lepton rapidities will enhance the contribution of higher-order radiative QCD effects and subleading production channels sensitive to the gluon PDF.
See e.g.~\cite{Hautmann:2012sh,Dooling:2012uw}. 
It is envisaged that studies of the FB asymmetry could be applied to investigate soft-gluon effects on the behaviour of PDFs~\cite{Hautmann:2017xtx,Hautmann:2017fcj}, and possibly set physical constraints
particularly in the large-$x$ region, which is still little constrained by experimental data, and where results for PDFs strongly depend at present on the choice of parameterizations used in global fits.

\section{Conclusions}
\label{sec:summa}
Accurate PDF sets are a precondition not only for refined tests of SM interactions, whether of QCD or EW origin, but also in the search for new physics phenomena.
A variety of experimental observables are used in order to constrain their evolution in $x$ and $Q^2$.
Amongst these, in hadronic collider environments, like those of Fermilab and LHC, DY processes, both via CCs and NCs, are those providing some of the strongest constraints.
A vast literature exists in this respect, largely concentrating on the scope offered by CC and NC differential cross section measurements in di-lepton mass and rapidity.
The CC asymmetry is widely used as well in PDF determinations. 
In this paper we have explored the case of the NC asymmetry, encoding information on the single lepton angular distribution.
This corresponds to triply differential cross sections in lepton pseudorapidity and di-lepton mass and rapidity.
We have highlighted that significant scope exists, in the pursuit of accurate PDF sets, from the exploitation of the NC process, as both its cross section and FB asymmetry are strongly sensitive to the underlying PDF dynamics.
In fact, the ability of fully reconstructing the final state (di-lepton) kinematics, which is not possible in the CC case, enables one to perform selected $Q^2$ fits, for both on-shell and off-shell $Z$-bosons,
both below and above $M_Z$, in regions in $x$ that are important for the aforementioned studies.
The case has been made here based on Run-I and Run-II data samples already collected or expected at the LHC, yet it becomes stronger as the accrued luminosity increases.
Indeed, HL-LHC luminosities may even afford one the potential of finely disentangling different PDF sets and corresponding parametrisations.

\section*{Acknowledgements}
\noindent
This work is supported by the Science and Technology Facilities Council (STFC), Grant No. ST/P000711/1, and by the BMBF under Contract No. 05H15PMCCA. 
F.H. acknowledges the support and hospitality of DESY, Hamburg and the University of Basque Country, Bilbao.
All authors acknowledge partial financial support through the NExT Institute. F.H. would like to thank the University of Hamburg and the University of Southampton for funding through a Staff Exchange Fellowship.
We thank Claire H. Shepherd-Themistocleous, Joey Huston and Valerio Bertone for useful discussions.

\bibliographystyle{apsrev4-1}
\bibliography{bib}

\end{document}